\documentclass[12pt]{article}
\pdfoutput=1

\usepackage{amsmath,hyperref}
\usepackage{fullpage}

\newtheorem{theorem}{Theorem}
\newtheorem{definition}{Definition}

\begin{document}

\title{A sufficient condition for counterexamples\\
       to the Nelson-Seiberg theorem}
\author{Zheng Sun\textsuperscript{*}, Zipeng Tan\textsuperscript{\dag}, Lu Yang\textsuperscript{\ddag}\\
        \normalsize\textit{College of Physics, Sichuan University,}\\
        \normalsize\textit{29 Wangjiang Road, Chengdu 610064, P.~R.~China}\\
        \normalsize\textit{E-mail:}
        \textsuperscript{*}\texttt{sun\_ctp@scu.edu.cn,}
        \textsuperscript{\dag}\texttt{tzpcyc@126.com,}
        \textsuperscript{\ddag}\texttt{1546265328@qq.com}
       }
\date{}
\maketitle

\begin{abstract}
Several counterexample models to the Nelson-Seiberg theorem have been discovered in previous literature, with generic superpotentials respecting the R-symmetry and non-generic R-charge assignments for chiral fields.  This work present a sufficient condition for such counterexample models:  The number of R-charge $2$ fields, which is greater than the number of R-charge $0$ fields, must be less than or equal to the number of R-charge $0$ fields plus the number of independent field pairs with opposite R-charges and satisfying some extra requirements.  We give a correct count of such field pairs when there are multiple field pairs with degenerated R-charges.  These models give supersymmetric vacua with spontaneous R-symmetry breaking, thus are counterexamples to both the Nelson-Seiberg theorem and its extensions.
\end{abstract}

\section{Introduction}

The Nelson-Seiberg theorem~\cite{Nelson:1993nf} and its extensions~\cite{Kang:2012fn, Li:2020wdk} provide a method using R-symmetries to build F-term supersymmetry (SUSY) breaking vacua~\cite{Intriligator:2007cp} in the framework of Wess-Zumino models~\cite{Wess:1973kz, Wess:1974jb} or O'Raifeartaigh models~\cite{ORaifeartaigh:1975nky}.  The theorem also applies to metastable SUSY breaking models~\cite{Intriligator:2006dd} with approximate R-symmetries~\cite{Intriligator:2007py, Abe:2007ax}.  SUSY preserving vacua can also be build with properly arranged R-charges of fields~\cite{Sun:2011fq, Brister:2021xxxx}.  These vacua appear in the model building of both SUSY phenomenology beyond the Standard Model~\cite{Nilles:1983ge, Martin:1997ns, Baer:2006rs, Terning:2006bq, Dine:2007zp} and compactification of string theory~\cite{Grana:2005jc, Douglas:2006es, Blumenhagen:2006ci, Ibanez:2012zz, Blumenhagen:2013fgp}.

The proof of both the Nelson-Seiberg theorem and its extensions require generic superpotentials.  The term genericness usually refers to generic values of the superpotential parameters.  Non-generic models of this type require fine-tuning of parameters, and are not interesting for phenomenology studies.  Another type of non-genericness comes from the specific form of the superpotential restricted by the R-symmetry with some special R-charge arrangement.  Such non-generic models do not suffer from parameter tuning, and can be referred to as counterexamples to the Nelson-Seiberg theorem with non-generic R-charges.  The first counterexample model of this type has been discovered recently~\cite{Sun:2019bnd} and thereafter generalized~\cite{Amariti:2020lvx}.  A common feature of these models is that they have at least one field pairs with opposite R-charges.  The product of such field pairs act effectively like R-charge $0$ fields for the field counts in the revised Nelson-Seiberg theorem~\cite{Kang:2012fn}, so SUSY vacua exist in these models even with more R-charge $2$ fields than R-charge $0$ fields.  Following this idea, a sufficient condition for counterexamples is presented in~\cite{Amariti:2020lvx}, and new counterexamples are accordingly constructed using multiple field pairs with opposite R-charges.

In this work, we are to present a more general sufficient condition for counterexamples.  In addition to the one in~\cite{Amariti:2020lvx}, we consider the case with multiple R-charge $2$ and R-charge $0$ fields.  We also give a correct count of independent field pairs with opposite R-charges when there are multiple field pairs with degenerated R-charges.  The sufficient condition is stated in terms of these field counts.  The statement is then supported with examples which give SUSY vacua with spontaneous R-symmetry breaking, as predicted by field counts satisfying the sufficient condition.

The rest part of this paper is arranged as following.  Section 2 reviews the Nelson-Seiberg theorem, its extensions and previously found counterexamples.  Section 3 presents the sufficient condition with a proof, and discusses properties of vacua from counterexamples.  Section 4 gives examples supporting our statement.  Section 5 makes the conclusion and final remarks.

\section{The Nelson-Seiberg theorem and counterexamples}

In a Wess-Zumino model, SUSY vacua are solutions to the F-term equations
\begin{equation}
\partial_i W = \frac{\partial W}{\partial \phi_i}
             = 0, \label{eq:2-01}
\end{equation}
where $W = W(\phi_i)$, named the superpotential, is a holomorphic function of chiral superfields $\phi_i$ for $i = 1, \dotsc, n$.  Such a solution also corresponds to a minimum of the scalar potential
\begin{equation}
V = K^{\bar{i} j} (\partial_i W)^* \partial_j W
\end{equation}
with the vacuum expectation value (VEV) $\langle V \rangle = 0$.  Einstein summation is used in the expression of $V$, and the K\"ahler metric $K^{\bar{i} j}$ is calculated from a K\"ahler potential $K(\phi_i^*, \phi_j)$ which is a real and positive-definite function of $\phi_i$'s and their conjugates $\phi_i^*$'s.  The vacuum is determined just by the scalar components $z_i$'s of $\phi_i$'s, so $W$ and $K$ are also viewed as functions of $z_i$'s.  If~\eqref{eq:2-01} has no solution, SUSY must be broken at any vacuum of the model, although we need to assume the existence of a local minimum of $V$, or build a minimum by introducing various corrections to a runaway direction~\cite{Ferretti:2007ec, Ferretti:2007rq, Azeyanagi:2012pc, Sun:2018hnk}.  Taking the non-existence of a solution to~\eqref{eq:2-01} as the criteria for SUSY breaking, the Nelson-Seiberg theorem and its extensions~\cite{Nelson:1993nf, Kang:2012fn, Li:2020wdk} give the conditions for SUSY breaking, expressed in terms of R-symmetries under which $W$ has R-charge $2$.  These theorems are as follows:

\begin{theorem}
(The Nelson-Seiberg theorem)
In a Wess-Zumino model with a generic superpotential, assuming the existence of a vacuum at the global minimum of the scalar potential, an R-symmetry is a necessary condition, and a spontaneously broken R-symmetry is a sufficient condition for SUSY breaking at the vacuum.
\end{theorem}

\begin{theorem}
(The Nelson-Seiberg theorem revised and generalized)
In a Wess-Zumino model with a generic superpotential, assuming the existence of a vacuum at the global minimum of the scalar potential, SUSY is spontaneously broken at the vacuum if and only if the superpotential has an R-symmetry, and one of the following conditions is satisfied:
\begin{itemize}
\item The superpotential is smooth at the origin of the field space, and the number of R-charge $2$ fields is greater than the number of R-charge $0$ fields for any possible consistent R-charge assignment.
\item The superpotential is singular at the origin of the field space.
\end{itemize}
\end{theorem}

For a generic R-symmetric polynomial $W$, the criteria for SUSY breaking is just the comparison between R-charge $2$ and R-charge $0$ field counts, which can be done much more easily and quickly than solving the F-term equations~\eqref{eq:2-01}.  Thus it is possible to do a fast scan of a large number of models based on the field counting criteria.  The accuracy of such a scan is affected by the existence of counterexamples with generic parameters and non-generic R-charges.  The simplest counterexample model~\cite{Sun:2019bnd} of this type has four fields $\{z_1, z_2, z_3, z_4\}$ with R-charges
\begin{equation}
(R(z_1), R(z_2), R(z_3), R(z_4)) = (2, -2, 6, -6), \label{eq:2-02}
\end{equation}
and the superpotential
\begin{equation}
W = a z_1 + b z_1^2 z_2 + c z_2^2 z_3 + d z_1 z_3 z_4 \label{eq:2-03}
\end{equation}
which includes all renormalizable terms of R-charge $2$.  The R-charge assignment is also uniquely fixed by requiring all terms of $W$ to have R-charge $2$.  The model has one R-charge $2$ field $z_1$ and no R-charge $0$ field, so the field counting criteria predicts SUSY breaking.  But the F-term equations have a SUSY solution
\begin{equation}
z_1 = z_2
    = 0, \quad
z_3 z_4 = - \frac{a}{d},
\end{equation}
which also breaks the R-symmetry for generic non-zero values of parameters $a$ and $d$.  Thus the model is a counterexample to both the original and the revised Nelson-Seiberg theorems.  The counterexample does not falsify either theorem, since the particular R-charge assignment restrict $W$ to particular form which violates the genericness assumption of the theorems.  It is significative to find out the pattern of R-charges in such models rather than simply identify them as counterexamples with non-generic R-charges.

One feature of the R-charge assignment~\eqref{eq:2-02} is the existence of the oppositely R-charged field pair $z_3$ and $z_4$.  They appear as a product $z_3 z_4$ in the superpotential~\eqref{eq:2-03}, although $z_3$ also appears linearly in another cubic term.  The work of~\cite{Amariti:2020lvx} explores this feature and presents a sufficient condition for counterexamples.  New counterexamples are constructed using one R-charge $2$ field, no R-charge $0$ field, and more than one field pairs with opposite R-charges.  At least one pair of oppositely R-charged fields satisfy the condition that they both appear only linearly in the superpotential, and are not involved in any quadratic term.  The product of such a field pair gets a non-zero VEV, and acts effectively like an R-charge $0$ field to help solving the F-term equations.  It is natural to generalize the sufficient condition and the counterexample construction to the case with multiple R-charge $2$ and R-charge $0$ fields.  Such generalization is to be done in the following section.

\section{A sufficient condition for counterexamples}

Following the convention in~\cite{Sun:2011fq} which is also used in~\cite{Kang:2012fn, Li:2020wdk}, we make the following field classification in an R-symmetric Wess-Zumino model:
\begin{definition} \label{def:01}
(The field classification)
Under an R-symmetry, fields are classified according to their R-charges into the following types:
\begin{itemize}
\item $X$ fields:  $X_i$ for $i = 1, \dotsc, N_X$ with $R(X_i) = 2$;
\item $Y$ fields:  $Y_i$ for $i = 1, \dotsc, N_Y$ with $R(Y_i) = 0$;
\item $P_{(r)}$ and $Q_{(- r)}$ fields for a value of $r$:  $P_{(r) i}$ for $i = 1, \dotsc, N_{P(r)}$ and $Q_{(- r) j}$ for $j = 1, \dotsc, N_{Q(- r)}$ with $R(P_{(r) i}) = - R(Q_{(- r) j}) = r$, and both $P$'s and $Q$'s only appear linearly in the superpotential and not in any quadratic terms;
\item $A$ fields:  $A_i$ for $i = 1, \dotsc, N_A$ with $R(A_i) = r_i \ne 2$ or $0$, and $A$'s do not satisfy the condition for being classified as $P_{(r)}$ and $Q_{(- r)}$ fields.
\end{itemize}
\end{definition}
For a renormalizable superpotential, both $P$'s and $Q$'s only appear linearly in cubic terms.  Other oppositely R-charged field pairs which can not be classified as $P$-$Q$ pairs are identified as $A$'s.  This field classification leads to the generic, R-symmetric and renormalizable superpotential
\begin{equation}
\begin{aligned}
W &= W_0 + W_1,\\
W_0 &= X_i (a_i + b_{i j} Y_j + c_{i j k} Y_j Y_k
            + d_{(r) i j k} P_{(r) j} Q_{(- r) k}),\\
W_1 &= \underbrace{\xi_{i j} X_i^2 A_j}_{r_j = - 2}
       + \underbrace{\rho_{i j k} X_i A_j A_k}_{r_j + r_k = 0}
       + \underbrace{\sigma_{(r) i j k} P_{(r) i} A_j A_k}_{r_j + r_k = 2 - r}
       + \underbrace{\tau_{(r) i j k} Q_{(- r) i} A_j A_k}_{r_j + r_k = 2 + r}\\
    &\quad
       + \underbrace{(\mu_{i j} + \nu_{i j k} Y_k) A_i A_j}_{r_i + r_j = 2}
       + \underbrace{\lambda_{i j k} A_i A_j A_k}_{r_i + r_j + r_k = 2}.
\end{aligned} \label{eq:3-01}
\end{equation}
All terms of $W_1$ are at least quadratic in $X$'s and $A$'s.  This statement is also true for non-renormalizable superpotentials.  So if we look for a vacuum satisfying
\begin{equation}
X_i = A_j
    = 0,
\end{equation}
the first derivatives of $W_0$ respecting to $Y$'s, $P$'s, $Q$'s and $A$'s, as well as all first derivatives of $W_1$ vanish at such a vacuum.  The F-term equations are then reduced to
\begin{equation}
a_i + b_{i j} Y_j + c_{i j k} Y_j Y_k + d_{(r) i j k} P_{(r) j} Q_{(- r) k} = 0, \label{eq:3-02}
\end{equation}
or more generally
\begin{equation}
f_i(Y_j, P_{(r) k} Q_{(- r) l}) = 0 \label{eq:3-03}
\end{equation}
which also covers the non-renormalizable case.  Suppose that the total number of independent quadratic products $P_{(r) j} Q_{(- r) k}$, or the number of independent $P$-$Q$ pairs with opposite R-charges, is $N_{P Q}$.  These quadratic products can be effectively viewed as $N_{P Q}$ independent variables.  So there are totally $N_Y + N_{P Q}$ variables to solve $N_X$ equations.  SUSY solutions exist with $N_X \le N_Y + N_{P Q}$ and generic parameters.  But the revised theorem predicts SUSY breaking when $N_X > N_Y$.  Notice also that a solution to~\eqref{eq:3-02} or~\eqref{eq:3-03} generically gives non-zero VEV's for $P$'s and $Q$'s, which spontaneously break the R-symmetry.  Thus the generic superpotential~\eqref{eq:3-01} under the condition
\begin{equation}
N_Y < N_X
    \le N_Y + N_{P Q}
\end{equation}
gives a counterexample to both the original Nelson-Seiberg theorem and the revised one.

The total number of $P$-$Q$ pairs is
\begin{equation}
N'_{P Q} = \sum_r N_{P(r)} N_{Q(- r)}.
\end{equation}
But the corresponding quadratic products may not be independent when there are multiple fields with degenerated R-charges.  The set
\begin{equation}
\bigcup_r \{ P_{(r) j} Q_{(- r) k} \mid j = 1 \lor k = 1 \} \label{eq:3-04}
\end{equation}
contains independent elements, since every quadratic product in this set except $P_{(r) 1} Q_{(- r) 1}$ has either $P_{(r) j}$ or $Q_{(- r) k}$ not appearing in other set elements.  Quadratic products not in~\eqref{eq:3-04}, if existing, can be expressed in terms of the elements of~\eqref{eq:3-04}:
\begin{equation}
P_{(r) j} Q_{(- r) k} = \frac{\left ( P_{(r) j} Q_{(- r) 1} \right ) \left ( P_{(r) 1} Q_{(- r) k} \right )}{P_{(r) 1} Q_{(- r) 1}}, \quad
j, k > 1.
\end{equation}
Thus the order of the set~\eqref{eq:3-04} gives the number of independent $P$-$Q$ pairs
\begin{equation}
N_{P Q} = \sum_r \left ( N_{P(r)} + N_{Q(- r)} - 1 \right ). \label{eq:3-05}
\end{equation}
Notice that $N_{P Q} < N'_{P Q}$ if there is at least one $r$ with $N_{P(r)} > 1$ and $N_{Q(- r)} > 1$.

In summary, we have obtained the sufficient condition:

\begin{theorem} \label{thm:03}
(A sufficient condition for counterexamples to the Nelson-Seiberg theorem)
In a Wess-Zumino model with a generic R-symmetric superpotential, using the notation in Definition~\ref{def:01}, the condition $N_Y < N_X \le N_Y + N_{P Q}$ is a sufficient condition for the model to be a counterexample to both the original Nelson-Seiberg theorem and the revised one, where $N_{P Q} = \sum_r \left ( N_{P(r)} + N_{Q(- r)} - 1 \right )$ is the number of independent $P$-$Q$ pairs.
\end{theorem}

The total number of $P$'s and $Q$'s is
\begin{equation}
N_{P + Q} = \sum_r \left ( N_{P(r)} + N_{Q(- r)} \right ).
\end{equation}
The reduced F-term equations~\eqref{eq:3-02} or~\eqref{eq:3-03} can also be viewed as using $N_Y + N_{P + Q}$ variables to solve $N_X$ equations.  Generically each solution has $N_Y + N_{P + Q} - N_X$ undetermined variables, or degeneracy of complex dimension $N_Y + N_{P + Q} - N_X$.  Since $N_{P + Q}$ is greater than $N_{P Q}$ for models with at least one $P$-$Q$ pairs, SUSY solutions in counterexamples have degeneracy dimension of at least one.  The VEV's of $P$'s and $Q$'s are generically non-zero, and the R-symmetry is spontaneously broken everywhere on the degenerated vacuum.  Such a SUSY solution also makes the superpotential~\eqref{eq:3-01} to be zero, thus the bound on the superpotential is satisfied~\cite{Kappl:2008ie, Dine:2009sw}.  These properties of vacua from counterexample models are summarized in the following theorem:

\begin{theorem} \label{thm:04}
(Properties of vacua from counterexamples)
A counterexample model according to Theorem~\ref{thm:03} gives one or several SUSY vacua with $W = 0$ and degeneracy of complex dimension $N_Y + N_{P + Q} - N_X$, where $N_{P + Q} = \sum_r \left ( N_{P(r)} + N_{Q(- r)} \right )$ is the total number of $P$'s and $Q$'s.  The R-symmetry is spontaneously broken by the non-zero VEV's of $P$-$Q$ pairs everywhere on the degenerated vacua.
\end{theorem}

\section{Examples of counterexamples}

The sufficient condition is demonstrated by the following examples of counterexample models.  The simplest counterexample~\eqref{eq:2-02} and~\eqref{eq:2-03} can be rewritten using the notation in in Definition~\ref{def:01}, as the R-charge assignment
\begin{equation}
(R(X), R(P), R(Q), R(A)) = (2, 6, -6, -2)
\end{equation}
and the superpotential
\begin{equation}
W = X (a + d P Q) + \xi X^2 A + \sigma P A^2.
\end{equation}
It has $N_X = N_{P(6)} = N_{Q(- 6)} = 1$, $N_Y = 0$, and thus $N_{P Q} = 1$.  The quadratic product $P Q$, viewed effectively as an R-charge $0$ variable, solves the reduced F-term equation $a + d P Q = 0$.  The SUSY solution is given as
\begin{equation}
X = A
  = 0, \quad
P Q = - \frac{a}{d},
\end{equation}
and the R-symmetry is broken by the non-zero VEV's of $P$ and $Q$.

As an example of models with both $Y$'s and $P$-$Q$ pairs, the R-charge assignment
\begin{equation}
(R(X_1), R(X_2), R(Y), R(P), R(Q), R(A)) = (2, 2, 0, 6, -6, -2)
\end{equation}
leads to the superpotential
\begin{equation}
\begin{split}
W &= X_1 (a_1 + b_1 Y + c_1 Y^2 + d_1 P Q)
     + X_2 (a_2 + b_2 Y + c_2 Y^2 + d_2 P Q)\\
  &\quad  
     + \xi_1 X_1^2 A
     + \xi_2 X_2^2 A
     + \sigma P A^2.
\end{split}
\end{equation}
This model has $N_X = 2$, $N_Y = N_{P(6)} = N_{Q(- 6)} = 1$, and thus $N_{P Q} = 1$.  The two independent effective variables $P Q$ and $Y$ solve the two reduced F-term equations.  The SUSY solution is given as
\begin{equation}
\begin{gathered}
X_1 = X_2
    = A
    = 0, \quad
Y = \frac{- b_1 d_2 + b_2 d_1 \pm \sqrt{\Delta}}{2 (c_1 d_2 - c_2 d_1)},\\
P Q = \frac{a_1 c_2 - a_2 c_1}{c_1 d_2 - c_2 d_1}
      + \frac{(b_1 c_2 - b_2 c_1)(- b_1 d_2 + b_2 d_1 \pm \sqrt{\Delta})}{2 (c_1 d_2 - c_2 d_1)^2},\\
\text{where} \
\Delta = (b_1 d_2 - b_2 d_1)^2 - 4 (a_1 d_2 - a_2 d_1) (c_1 d_2 - c_2 d_1),
\end{gathered}
\end{equation}
and the R-symmetry is broken by the non-zero VEV's of $P$ and $Q$.

As an example of models with some oppositely R-charged field pairs which can not be classified as $P$-$Q$ pairs, the R-charge assignment
\begin{equation}
(R(X), R(P), R(Q), R(A_1), R(A_2)) = (2, 4, -4, 1, -1)
\end{equation}
leads to the superpotential
\begin{equation}
W = X (a + d P Q) + \rho X A_1 A_2 + \sigma P A_2^2 + \mu A_1^2.
\end{equation}
The $A_1$-$A_2$ pair, although with opposite R-charges, can not be classified as $P$'s and $Q$'s, because both $A_1$ and $A_2$ appear quadraticly in $W$, and $A_1$ appears in a quadratic term.
This model has $N_X = N_{P(4)} = N_{Q(- 4)} = 1$, $N_Y = 0$, and thus $N_{P Q} = 1$.  The SUSY solution is given as
\begin{equation}
X = A_1
  = A_2
  = 0, \quad
P Q = - \frac{a}{d},
\end{equation}
and the R-symmetry is broken by the non-zero VEV's of $P$ and $Q$.

As an example of models which contain multiple $P$-$Q$ pairs with degenerated R-charges, the R-charge assignment
\begin{equation}
(R(X_1), R(X_2), R(X_3), R(P_1), R(P_2), R(Q_1), R(Q_2), R(A))
= (2, 2, 2, 6, 6, -6, -6, -2)
\end{equation}
leads to the superpotential
\begin{equation}
\begin{split}
W &= X_1 (a_1
          + d_{1 1 1} P_1 Q_1
          + d_{1 1 2} P_1 Q_2
          + d_{1 2 1} P_2 Q_1
          + d_{1 2 2} P_2 Q_2)\\
  &\quad
     + X_2 (a_2
            + d_{2 1 1} P_1 Q_1
            + d_{2 1 2} P_1 Q_2
            + d_{2 2 1} P_2 Q_1
            + d_{2 2 2} P_2 Q_2)\\
  &\quad
     + X_3 (a_3
            + d_{3 1 1} P_1 Q_1
            + d_{3 1 2} P_1 Q_2
            + d_{3 2 1} P_2 Q_1
            + d_{3 2 2} P_2 Q_2)\\
  &\quad
     + \xi_1 X_1^2 A
     + \xi_2 X_2^2 A
     + \xi_3 X_3^2 A
     + \sigma_1 P_1 A^2
     + \sigma_2 P_2 A^2.
\end{split}
\end{equation}
This model has $N_X = 3$, $N_Y = 0$, $N_{P(6)} = N_{Q(- 6)} = 2$, and thus
\begin{equation}
N_{P Q} = N_{P(6)} + N_{Q(- 6)} - 1
        = 3.
\end{equation}
SUSY solutions correspond to
\begin{equation}
X_1 = X_2
    = X_3
    = A
    = 0 \label{eq:4-01}
\end{equation}
and the solutions to
\begin{equation}
\begin{aligned}
a_1
+ d_{1 1 1} P_1 Q_1
+ d_{1 1 2} P_1 Q_2
+ d_{1 2 1} P_2 Q_1
+ d_{1 2 2} P_2 Q_2 & = 0,\\
a_2
+ d_{2 1 1} P_1 Q_1
+ d_{2 1 2} P_1 Q_2
+ d_{2 2 1} P_2 Q_1
+ d_{2 2 2} P_2 Q_2 & = 0,\\
a_3
+ d_{3 1 1} P_1 Q_1
+ d_{3 1 2} P_1 Q_2
+ d_{3 2 1} P_2 Q_1
+ d_{3 2 2} P_2 Q_2 & = 0.
\end{aligned} \label{eq:4-02}
\end{equation}
The three independent effective variables $P_1 Q_1$, $P_1 Q_2$ and $P_2 Q_1$ solve the three reduced F-term equations.  Then $P_1$, $P_2$, $Q_1$ and $Q_2$ can be expressed in terms of these three variables and a free parameter representing the degeneracy.  The analytical solution is a radical expression of several hundred terms, which is too complicated to be presented here.  Numerical solutions with some typical choices of coefficient values are listed in Table~\ref{tb:01}.  These solutions, together with~\eqref{eq:4-01}, give SUSY vacua with R-symmetry breaking by the non-zero VEV's of $P_1$, $P_2$, $Q_1$ and $Q_2$.

\begin{table}
    \centering
    \begin{tabular}{|c|c|c|}
        \hline
        $(a_1, d_{1 1 1}, d_{1 1 2}, d_{1 2 1}, d_{1 2 2})$ &
        $(a_2, d_{2 1 1}, d_{2 1 2}, d_{2 2 1}, d_{2 2 2})$ &
        $(a_3, d_{3 1 1}, d_{3 1 2}, d_{3 2 1}, d_{3 2 2})$\\
        \hline \hline
        $(1.2, - 1.3, 1.8, 1.3, - 0.8)$ &
        $(- 0.6, 0.9, 1.2, 1.8, 0.7)$ &
        $(- 1.9, - 1.6, - 1.1, - 1.2, 1.1)$\\
        \hline
        \multicolumn{3}{|c|}{$P_1 Q_1 = - 0.267532, \ P_2 = - 1.73027 P_1, \ Q_2 = 2.52332 Q_1$}\\
        \multicolumn{3}{|c|}{or $P_1 Q_1 = - 0.0503819, \ P_2 = 7.04817 P_1, \ Q_2 = - 4.15665 Q_1$}\\
        \hline \hline
        $(- 1.7, - 1.1, 1.5, - 0.8, 1.4)$ &
        $(- 1.2, 1.7, 0.5, 1.9, 0.6)$ &
        $(1.7, - 1.9, - 1.5, - 0.6, 0.7)$\\
        \hline
        \multicolumn{3}{|c|}{$P_1 Q_1 = - 5.96733, \ P_2 = - 1.04776 P_1, \ Q_2 = - 0.696789 Q_1$}\\
        \multicolumn{3}{|c|}{or $P_1 Q_1 = 0.208592, \ P_2 = 0.345409 P_1, \ Q_2 = 4.80255 Q_1$}\\
        \hline \hline
        $(0.8, - 1.6, 0.7, - 0.6, - 1.8)$ &
        $(- 1.6, - 1.5, - 0.5, - 1.7, 0.9)$ &
        $(- 1.7, 0.9, - 0.6, 1.9, - 1.7)$\\
        \hline
        \multicolumn{3}{|c|}{$P_1 Q_1 = - 3.50319, \ P_2 = - 1.01492 P_1, \ Q_2 = 0.48258 Q_1$}\\
        \multicolumn{3}{|c|}{or $P_1 Q_1 = - 0.320605, \ P_2 = 0.106889 P_1, \ Q_2 = 8.19428 Q_1$}\\
        \hline \hline
        $(- 0.6, 0.8, - 1.3, - 1.4, - 1.1)$ &
        $(- 1.6, - 1.3, - 1.4, 1.8, - 0.7)$ &
        $(1.3, - 0.6, - 1.7, 1.7, 0.6)$\\
        \hline
        \multicolumn{3}{|c|}{$P_1 Q_1 = - 4.32013, \ P_2 = 0.581242 P_1, \ Q_2 = 0.064529 Q_1$}\\
        \multicolumn{3}{|c|}{or $P_1 Q_1 = - 0.14664, \ P_2 = - 3.28734 P_1, \ Q_2 = - 4.09916 Q_1$}\\
        \hline
    \end{tabular}
    \caption{Numerical solutions to~\eqref{eq:4-02} with typical choices of coefficient values.} \label{tb:01}
\end{table}

All models in this section satisfy $N_Y < N_X \le N_Y + N_{P Q}$, thus are counterexamples to the Nelson-Seiberg theorem according to Theorem~\ref{thm:03}.  The analytically or numerically obtained SUSY solutions have $W = 0$, degeneracy of $N_Y + N_{P + Q} - N_X$, and R-symmetry breaking everywhere on the degenerated vacua.  So both Theorem~\ref{thm:03} and Theorem~\ref{thm:04} are verified by these examples.

\section{Outlook}

In this work, we investigate features of counterexamples to the Nelson-Seiberg theorem, and successfully proved a theorem which provides a sufficient condition for counterexamples.  The scope of the theorem covers all previously found counterexamples in literature~\cite{Sun:2019bnd, Amariti:2020lvx}.  The sufficient condition is expressed as the comparison between field counts of different R-charges, and the count of independent $P$-$Q$ pairs has the simple expression~\eqref{eq:3-05}.  It is still feasible to do a fast survey of a large number of models using the field counting method, even taking into account the counterexamples in this work.  Thus the pattern of field counts in counterexamples enables a refined classification of R-symmetric Wess-Zumino models.

Counterexample models built from our sufficient condition have certain properties, which may inspire applications in both phenomenology and formal studies.  The R-symmetry breaking SUSY vacua are applicable to model building of tree-level R-symmetry breaking~\cite{Carpenter:2008wi, Sun:2008va, Komargodski:2009jf, Liu:2014ida}, where the R-symmetry breaking sector can be build separately from the SUSY breaking sector.  The SUSY vacua with $W = 0$ become Minkowski SUSY vacua in supergravity, which make up the low-energy SUSY branch of the string landscape~\cite{Dine:2004is, Dine:2005yq, Dine:2005gz}.  The degeneracy direction of vacua can be used for cosmological model building with non-perturbative potentials introduced, and the difficulty from the de Sitter swampland conjecture may be avoided~\cite{Obied:2018sgi, Garg:2018reu, Ooguri:2018wrx, Palti:2019pca}.  It is still challenging to find ultraviolet completion of these models in strongly coupled SUSY gauge theories or compactification of string theory.

\section*{Acknowledgement}

We thank Yan He and Jinmian Li for helpful discussions.  This work is supported by the National Natural Science Foundation of China under grant 11305110.

\end{document}